\documentclass [12pt,a4paper,draft]{article}
\usepackage{times}

\DeclareFontFamily{OT1}{times}{}
\DeclareFontShape {OT1}{times}{m }{n }{ <-> ptmr }{}
\DeclareFontShape {OT1}{times}{bx}{n }{ <-> ptmb }{}
\DeclareFontShape {OT1}{times}{m }{it}{ <-> ptmri}{}
\DeclareFontShape {OT1}{times}{bx}{it}{ <-> ptmbi}{}

\usepackage{amsmath}
\usepackage{amsfonts}
\usepackage{amssymb}
\usepackage{latexsym}
\begin{document}

\title{Cornelius Lanczos --- Discoveries in the Quantum and General Relativity Theories}

\author{Mendel Sachs\\
Department of Physics, University at Buffalo,\\
State University of New York}

\date{To appear in {\bf Annales Fond. Louis de Broglie} (2002)}

\maketitle

\begin{abstract}

   Two of Lanczos' seminal contributions in physics that are not generally known about in the physics community are discussed and commented in relation to more recent investigations: First, his integral formulation of the Schr\"odinger equation that he published just before Schr\"odinger himself published his differential equation formulation of quantum mechanics, and, second, his solution to the problem of equations of motion of material particles in the theory of general relativity.

\end{abstract}

Cornelius Lanczos was indeed one of the greatly
inspired scholars in mathematical physics and
applied mathematics in the twentieth century.
It is not only his voluminous outpouring of
excellent research on diverse subjects in these
fields; it was perhaps more importantly his
open spirit that must influence his peers and
generations of physicists and mathematicians.
Thus, it is highly commendable that Lanczos'
collected papers and commentaries on them have
been published in \emph{Cornelius Lanczos: Collected
Published Papers with Commentaries}, editors:
W.R. Davis, M. Chu, P. Dolan, J. R. McConnell,
L. K. Norris, E. Ortiz, R. J. Plemmons,
D. Ridgeway, B. K. P. Scaife, W. J. Stewart,
J. W. York, Jr., W. O. Doggett, B. M. Gellai,
A. A. Gsponer, C. A. Prioli (North Carolina
State University, Raleigh, 1998), see [1]. Henceforth
this collection will be referred to as CLCPPC.

Because of our mutual research interests in the quantum
and general relativity theories, and questions concerned
with (nonrelativistic and relativistic) quantum theory,
I started a correspondence with Cornelius, when he was
in Dublin, in the 1960s. He kindly invited me to spend
some time with him at the Dublin Institute for Advanced
Studies. I enjoyed the opportunity to visit (in 1964 and
1973), to discuss questions of interest of both of us and
to have discussions with his colleagues at the Institute,
J. Synge, J. McConnell and L. O'Raifertaigh.

   In this note, I would like to briefly discuss two of
Lanczos' investigations that are not generally known
about in the physics community and to comment on them.

\section{A Discovery in the Quantum Theory}

In 1926, at the onset of the development of the formal expressions of
quantum mechanics, Lanczos discovered that Heisenberg's (discrete)
matrix representation and Schrödinger's (continuous) wave representation
are mathematically equivalent --- since each can be transformed into the
other. [Schrödinger also discovered this fact independently, in the same
year.]

   What it was that Lanczos saw was that the equations of motion and the
quantum condition could be expressed in the form of integral equations.
Thus he said: ``A conception of the continuum exists side by side with
equal validity with the conception of the discrete, because there is a
unique relationship between them.''

   The form of the Schrödinger wave equation is:
$$
     H \Psi \equiv i \frac{\partial \Psi}{\partial t} = -\nabla^2 \Psi + V \Psi = E \Psi         \eqno(1)
$$
where $H$ is the Hamiltonian operator (as defined by Schrödinger in terms
of the operator equivalents of the momentum and position variables of a
particle of matter), and $E$ is the particle's  energy eigenvalue of $H$ (units chosen above with $h/2\pi = 1$, where $h$ is Planck's constant).

   To derive the integral form of this equation, Lanczos considered the
solution of the mathematical problem
$$
      -\nabla^2 \Psi + V \Psi = u  ~~~ ,  ~~~  \Psi{(\text{boundary})} = 0
$$
He found that the boundary value problem may be solved by means of the
Green's function $K(P,Q)$ for this inhomogeneous differential equation,
with the solution
$$
    \Psi(P)  \equiv   \int K(P,Q) ~ u(Q) ~ dQ
$$
Replacing $u(Q)$ with $E\Psi(Q)$, and dividing by $E$, one has the integral
equation:
$$
      \int K(P,Q) ~ \Psi(Q) ~ dQ  =  (1/E) \Psi(P)    \eqno(2)
$$
Thus, the eigenvalues of the kernel $K(P,Q)$ are the reciprocals of the
energy values, $1/E$.

   The correspondence between equations (1) and (2) then leads to the
Heisenberg equations of motion for quantum mechanics,
$$
  [H,O] \Psi \equiv (HO - OH) \Psi =  i\frac{\partial O}{\partial t} \Psi     \eqno(3)
$$

This equation, in turn, yields the matrix representation of quantum
mechanics, where $H$ is the Hamiltonian operator for the dynamical system
and $O$ is an operator that corresponds with some particular observable
physical property of the microsystem.

   Thus Lanczos showed, unequivocally, that Heisenberg's equation of
motion for quantum mechanics, (3), is equivalent to Schrödinger's wave
mechanical form (1). The details of this correspondence are spelled out
in Lanczos' paper: ``Uber eine maissige Darstellung der neuen
Quantenmechanik'' (``On a Field Theoretical Representation of the New
Quantum Mechanics''), \emph{Zeitschrift fur Physik} {\bf 35}, 812 (1926). Its English translation is in CLCPPC, Vol. III, p. 2-858.

   There has been a controversy on the question of who first discovered
this equivalence of the Heisenberg and Schrödinger representations of
quantum mechanics --- was it Schrödinger, Pauli or Lanczos? (from their
respective points of view). There is an interesting dialogue and
discussion on this by B. L. van der Waerden, ``From Matrix Mechanics and
Wave Mechanics to a Unified Quantum Mechanics'', CLCPPC, Vol. III,
2-896, including a letter written by Pauli to Jordan on this
subject. Another interesting article on this subject is  by J. R.
McConnell, ``Commentary on Lanczos' "On a Field Theoretical
Representation of the New Quantum Mechanics"'', CLCPPC, Vol. III, 2-950.
Van der Waerden, in his dialogues, favors Lanczos as the actual
discoverer of the equivalence of these two representations of quantum
mechanics.

   I might add the following comment: While the differential form of
Schröd\-ing\-er's wave mechanics (1) is equivalent to Lanczos' integral
equation (2) --- which in turn leads to the Heisenberg representation (3)
--- the latter integral equation form cannot be extended to an expression
of wave mechanics in a curved spacetime, as required in general
relativity. This is because the Green's function is \emph{defined} in terms of
a linear mathematical formalism at the outset. Thus, in the
(necessarily) nonlinear, curved spacetime of general relativity, a
Green's function for this problem does not exist. On the other hand,
Schrödinger's differential equation form (1) for wave mechanics can be
extended to the nonlinear, curved spacetime. This is achieved by going
(smoothly) to a spinor-quaternion formalism, where the solutions of the
equations are now \emph{spinor variables} in a curved spacetime rather than the
scalar functions of  Schrödinger's nonrelativistic wave mechanics. In
this extended formalism, ordinary derivatives are replaced with
covariant derivatives. [This extension is indicated by the irreducible
representations of the \emph{Einstein group} --- the symmetry group that
underlies the general covariance requirement of general relativity
theory.] The covariant derivative of a spinor field is the sum of the
ordinary derivative and a spin-affine connection term. The spinor
solutions themselves, for the matter fields, are then the basis
functions of quaternion differential operators. This generalized
nonlinear expression of wave mechanics in general relativity cannot then
be interpreted in terms of a probability calculus, since the latter, \emph{by
definition}, requires a linear calculus.

   This advantage of the differential equation form of quantum mechanics
over the integral equation form is a significant point since the
integral equation form of nonrelativistic quantum mechanics, which
Lanczos addresses in this problem, does not extend to the nonlinear
curved spacetime. But the latter extension is necessary since
nonrelativistic quantum mechanics is supposed to be not more than an
\emph{approximation} for a generally relativistic theory of matter in the
microscopic domain. The latter generalization of quantum mechanics to a
nonlinear form cannot then be interpreted as the Copenhagen school does
in terms of a probability calculus.

 Lanczos himself indicated in his writings since the 1920s that a nonlinear extension of wave mechanics must follow. [See, e.g. ``Dirac's wellenmechanische Theorie des Elektrons und ihre feldtheoretische Ausgestaltung'' (``Dirac's Wave Mechanical Theory of the Electron and its Field Theoretical Interpretation''), \emph{Physikalisch Zeitschrift} {\bf 31}, 120 (1930), English translation in CLCPPC, Vol. III, p. 2-1226, related commentary by André Gsponer and Jean-Pierre Hurni ``Lanczos--Einstein--Petiau: From Dirac's Equation to Non-Linear Wave Mechanics,'' CLCPPC, Vol. III, p. 2-1248.]

   I have spelled out the details of this extension of the formal
expression of quantum mechanics to general relativity in my books: M.
Sachs, \emph{General Relativity and Matter} (Reidel, 1982) and its sequel, M.
Sachs, \emph{Quantum Mechanics from General Relativity} (Reidel, 1986). I have
shown in the latter book that the formal expression of quantum
mechanics, in terms of the linear Hilbert space, emerges as a linear
approximation for a generally covariant, nonlinear field theory of the
inertia of matter.

\section{A Discovery in General Relativity}

A second very important discovery of Lanczos had to do with the problem
of equations of motion of material particles in the theory of general
relativity. Without resorting to approximation methods [as in A.
Einstein, L. Infeld and B. Hoffmann, ``Gravitational Equation and the
Problem of Motion'', \emph{Annals of Mathematics} {\bf 39}, 65 (1938)] Lanczos
discovered that the equation of motion of a gravitational body is
implicit in Einstein's tensor field equations themselves, that is,
without the need to add extra equations of motion to the field
equations.  This is in contrast with the standard electromagnetic field
theory. In the latter case, the field equations are in terms of
Maxwell's equations while the equations of motion of a charged body,
subjected to an electromagnetic field, must be added, in the form of the
Lorentz equation of motion.  The latter difference is due, in part, to
the fact that the Einstein field theory is explicitly nonlinear while
the Maxwell field theory is explicitly linear. Thus, the Einstein field
theory of gravitation is more complete than the Maxwell field theory of
electromagnetism. This is because Einstein's theory purports to be a
`closed form' theory of matter while the Maxwell theory is not so.

   In his paper, ``The Dynamics of a Particle in General Relativity'',
\emph{Physical Review} {\bf 59}, 813 (1941), duplicated in CLCPPC, Vol. IV,  2-1650,
Lanczos discusses this problem. He starts out by explaining that, in his
view, ``the moving force [that accelerates a material body] comes out in
terms of a volume integral, extended over the matter-occupied central
field of the particle.'' He points out that no matter how one might
modify the spacetime metric \emph{inside of the particle}, no motion is
predicted. [This is a difficulty emphasized by Einstein in his
correspondences with Lanczos.] Thus, one needs a force \emph{external} to the
body acted upon that would cause it to accelerate --- as one has with
Newton's second law of motion.
   What Lanczos did to overcome this difficulty was to show that ``the
volume integral of the moving force can be transformed into a boundary
integral, extended over the border of the particle, or any closed
surface that includes this particle.'' He then concluded that the motion
law is established rigorously, without the need for approximation
methods. The law that he derived had the form of Newton's second law of
motion, \emph{when one treats the static condition only} (that is, for a
particle initially at rest). Lanczos' law of motion in general
relativity then takes the form:
$$
      \frac{d^2 \xi^i}{dx_4^2}  =  -\Gamma^{(e)}_{44,i}           \eqno(4)
$$
where $i = 1, 2$ or 3 are the three spatial coordinates, $x_4$ is the time
coordinate , $\xi^i$ are the spatial coordinates of the particle and
$\Gamma^{(e)}_{44,i}$ are the $i$th derivatives of the `44' component of the
affine connection of the curved spacetime field \emph{external to the
particle}.

   It is important to note that the particle's mass $m$ does not appear in
this equation of motion of a body in a gravitational field. (It is a
generalization of Galileo's discovery that the gravitational
acceleration of a body is independent of its inertial mass.) Lanczos
calls this ``equivalent to the law of the geodesic line''.

   Lanczos' imposed ``static condition'' to derive the law of motion (4)
is perhaps too restrictive to conclude from it a general law of motion.
In my own analysis of this problem in general relativity, I have
concluded that there can be no discrete particle of matter in the first
place, in the context of this theory. The geodesics of the spacetime are
the solutions of the dynamical problem for a `field concentration' that
we identify empirically with a `thing' --- e.g. an electron, a planet or a
galaxy. The metric solutions of general relativity are the regular (i.e.
nonsingular and analytic \emph{everywhere}) functions of the space and time
coordinates. Thus, there is no `inside' and `outside' of a discrete
material particle, in the context of the continuous field theory
implicit in general relativity. What is ``seen'' as a particle is, rather,
a continuous (though peaked) mode of a (regular) matter field. Thus,
Lanczos had no need, in the framework of the theory of general
relativity, to transform the continuous volume integral --- a natural
expression of a mode of the continuum --- into a surface integral. For the
volume integral itself is intimately related to the entire continuum;
there is in reality no surface within the continuum, separated from it.
This conclusion of the holism (essential connectivity) of a material
system expresses the essence of Mach's principle, which I have found in
my studies is an implication of Einstein's theory of general relativity,
as a general theory of matter.

These conclusions are in accord with Lanczos' assertion that  (real) singularities are to be excluded from the solutions of the generally covariant field equations of matter. His position, rather, is that ``matter is no more a singularity of the field, but an eigensolution of the field equations.''  [``Die neue Feldtheorie Einsteins'' (The New Field Theory of Einstein),  \emph{Ergebnisse der exakten Naturwissenschaften} {\bf 10} (1931), 97--132. Duplicated and translated in CLCPPC, Vol. IV, p. 2-1443].

While I agree with the exclusion of singularities from the solutions of a generally covariant field theory of matter, it is my position that the eigenvalue structure of the field equations is an asymptotic, but \emph{not} an exact feature of the field laws. One may see this difference, for example, on the one hand in Lanczos' comments that it is ``all the same whether we work with the homogeneous equations and admit singularities, or with the inhomogeneous one excluding singularities'' (ibid., CLCPPC, Vol. IV, p. 2-1439). On the other hand, it is my view, that, physically, there are no meaningful homogeneous equations in the first place. It is because we start in general relativity with a closed system at the outset, wherein the left hand side of the field equations (where the field intensities appear) is a representation of the right hand side (sources), \emph{and vice versa}. Still, one may approximate the right hand side, \emph{under special circumstances}, to be asymptotically small, replacing these terms with zero for practical purposes of calculation.

The difference between Lanczos' and my position here is related to the features of nonlinear differential equations. The solutions of the corresponding inhomogeneous and homogeneous nonlinear equations do not go smoothly into one another, under any conditions. That is, it is my position that, even asymptotically, so long as the right hand side is close to, but not exactly equal to zero, the solutions of one of the respective nonhomogeneous or homogeneous differential equations have features, related to physical properties, that are not duplicated in the solutions of the other.

One may compare this, as Lanczos does, with Newton's theory of gravitation, wherein we have Laplace's equation  $\nabla^2 \varphi = 0$ outside of the boundary of a gravitational body, and Poisson's equation, $\nabla^2 \varphi = \rho$ inside of this boundary, where $\varphi$ is the gravitational potential of the body and $\rho$ is its density.  If these would be nonlinear equations, as they would in a curved spacetime, then the solutions of Poisson's equation would not be a linear superposition of the solutions of Laplace's equation. Indeed, in general relativity there could not be any discrete boundary; there is only a (variable) continuum in space and time. Only in the linear limit (of a flat spacetime), as in Newton's theory of gravitation, can one assume a discrete boundary for a quantity of matter.  But the latter limit is not true in an exact sense in general relativity theory.

Cornelius Lanczos was a good friend and colleague of mine. From our
personal correspondences and contacts I know that whatever criticism I
have mentioned in this note would have been accepted in the spirit of
the essential role of controversy in science, to ensure its progress.
For this attitude as well as his contributions, he was indeed one of the
great scientists of the twentieth century.

\end{document}